\documentclass[english]{emulateapj}
\usepackage[english]{babel}
\usepackage{amsmath,amsthm}
\usepackage{amsfonts}
\usepackage{graphicx}
\usepackage{float}
\usepackage{amssymb}
\usepackage[normalem]{ulem}

\def\apjl{ApJL }

\def\apjs{ApJS }

\def\aap{A\&A }

\begin{document}

\title{Optical to X-rays supernovae light curves following shock breakout through a thick wind}

\author{Gilad Svirski$^{1}$, Ehud Nakar$^{1}$ and Re'em Sari$^{2}$}
\affil{1. Raymond and Beverly Sackler School of Physics \&
Astronomy, Tel Aviv University, Tel Aviv 69978, Israel\\
2. Racah Institute for Physics, The Hebrew University, Jerusalem
91904, Israel\\}

\begin{abstract}
Recent supernovae (SNe) observations have motivated renewed interest
in SN shock breakouts from stars surrounded by thick winds. In such
events the interaction with the wind powers the observed luminosity,
and predictions include observable hard X-rays. Wind breakouts on
timescales of a day or longer are currently the most probable for
detection. Here we study the signal that follows such events. We
start from the breakout of the radiation mediated shock, finding
that the breakout temperature can vary significantly from one event
to another ($10^4 - 5 \times 10^6$ K) due to possible deviation from
thermal equilibrium. In general, events with longer breakout pulse
duration, $t_{bo}$, are softer. We follow the observed radiation
through the evolution of the collisionless shock which forms after
the breakout of the radiation mediated shock. We restrict
the study of the collisionless shock evolution to cases where the
breakout itself is in thermal equilibrium, peaking in optical/UV. In
these cases the post-breakout emission contains two spectral
components - soft (optical/UV) and hard (X-rays and possibly soft
$\gamma$-rays). Right after the breakout pulse X-rays are strongly
suppressed, and they carry only a small fraction of the total
luminosity. The hard component becomes harder and its luminosity
rises quickly afterwards, gaining dominance at $\sim 10-50
\,t_{bo}$. The ratio of the peak optical/UV to the peak X-ray
luminosity depends mostly on the breakout time. In early breakouts
($t_{bo} \lesssim 20$ d for typical parameters) they are comparable,
while in late breakouts ($t_{bo} \gtrsim 80$ d for typical
parameters) the X-rays becomes dominant only after the total
luminosity has dropped significantly. In terms of prospects for
X-ray and soft gamma-ray detections, it is best to observe 100-500
days after explosions with breakout timescales between a week and a
month.
\end{abstract}

\section{Introduction}
The breakout of a supernova (SN) shock through the stellar surface
has been an active field of analytic and numerical research for
several decades (e.g. \citealt{Colgate1974}, \citealt{Weaver1976},
\citealt{Falk1978}, \citealt{Klein_Chevalier1978},
\citealt{Imshennik+1981}, \citealt{Ensman_Burrows1992},
\citealt{Matzner_McKee1999}). Recent advancement of observational
facilities led to  the discovery of several shock breakout
candidates
\citep[e.g.,][]{Campana06,Soderberg08,Gezari08,Schawinski2008,Modjaz2009,Ofek+2010,Gezari10,Arcavi11}
which motivated a revisit of the theory of shock breakout
\citep[e.g.,][]{Chevalier_Fransson2008,Katz+2010,Piro+2010,Nakar_Sari2010,Murase+2011,Katz_RMS+2011,Rabinak_Waxman2011,Couch+2011,Nakar_Sari2011,Chevalier_Irwin2011,Balberg_Loab2011,Moriya_Tominaga2011}.

The main observational challenge posed by shock breakouts from
stellar surfaces is their short duration (no more than an hour even
in the case of a red supergiant). However, in some cases massive
mass loss prior to the SN explosion can extend the breakout signal,
facilitating its detection. \cite{Ofek+2010} and
\cite{Chevalier_Irwin2011} show that a breakout through a thick wind
may occasionally extend long enough to account for the complete SN
luminosity light curve, and suggest that at least some type IIn SNe
are in fact powered by such breakouts.

When a star goes through a SN explosion, a radiation mediated shock
(RMS) is driven through its envelope, accelerating through the sharp
density drop near the stellar edge \citep{Sakurai60}. If the star is
not surrounded by a thick wind then the shock breaks out of the
stellar surface, producing an intense short pulse of UV-X-ray
photons. If, on the other hand, the star is surrounded by a wind
with an optical depth between the stellar edge and the observer
$>c/v$, where c is the speed of light and $v$ is the shock speed,
then the shock continues into the wind without releasing photons to
the observer. At this point the shock starts decelerating and a
reverse-forward shock structure is formed. The reverse shock is
driven into the SN ejecta, which is characterized by a sharp rise in
the amount of energy that is carried by slower moving material. The
reverse shock is driving an increasing amount of energy into the
shocked region, which keeps accumulating until the wind optical
depth drops to $c/v$. Beyond this point the wind cannot contain
anymore the radiation mediated forward shock and the shock is
breaking out of the wind, releasing all its energy as an intense
pulse. This pulse can contain much more energy than a wind-less
breakout, over timescales that are much longer. The general
properties of a wind breakout were recently discussed by
\cite{Ofek+2010} and \cite{Chevalier_Irwin2011}, who find the pulse
luminosity and temperature assuming thermal equilibrium, and by
\cite{Balberg_Loab2011} who also show how the progenitor mass and
explosion energy can be constrained by the pulse observational
properties.

If the wind density does not fall abruptly, the energy behind the
shock keeps growing after breakout, and the characteristic evolution
of the resultant signal supplies an important observational probe
for the interaction of the forward shock with the wind.
\cite{Katz+2011} showed that if the breakout of the RMS occurs
within the wind, i.e. the matter consisting the breakout layer is
spread over a scale comparable to the breakout radius rather than
being concentrated within a thin layer, then a collisionless shock
is bound to develop and replace the RMS in accelerating the matter
ahead. This collisionless shock heats the immediate post-shock
electrons to $k_BT_e\sim60$ keV \citep{Katz+2011,Murase+2011}, where $T_e$ is the
electron temperature and $k_B$ is the Boltzmann constant. The
cooling processes of these electrons affect the observed temperature
(and possibly the luminosity evolution), which is further influenced
by the interplay between photons and electrons along the photons
diffusion path to the observer.

Here we examine the evolution of the luminosity and the observed
temperature at and following a wind shock breakout. We consider only
breakouts through a wind that is dense enough to produce a breakout
pulse lasting days or longer, which are easier to detect. We
restrict the discussion on the observed radiation during the
collisionless shock evolution to scenarios where the unshocked
electrons ahead of the shock and the radiation that diffuses through
them on its way to the observer, are in thermal equilibrium. This
condition implies shock velocities $\lesssim 10^4$ km/s, where the
exact limiting velocity can vary with the gas metallicity. Finally,
we consider only cases where the shocks are cooling fast during the
entire evolution, i.e., the shock heated plasma cool within a
dynamical time. This condition is satisfied for regular SNe
parameters.  We pay attention to hydrodynamic, diffusion and cooling
timescales involved, and show that the breakout timescale is the
most dominant parameter in determining the post-breakout evolution.
We map the different types of post breakout behaviors and describe
their light curves and spectra evolution.

As this work was near completion, \cite{Chevalier_Irwin2012}
submitted a paper, presenting some of the conclusions derived below
in the context of explaining the low X-ray luminosity of SN 2006gy,
observed by \emph{Chandra} at an age of 3-4 months. The low
luminosity is explained as the result of three factors: two, which
are discussed in details below, are the domination of inverse
Compton over free-free emission in cooling the hot post-shock
electrons, and the energy degradation of the hard free-free emitted
photons over the diffusion path. The third is photoabsorption of
soft X-rays by partially ionized pre-shock gas. This latter process
is not considered in this work.

\section{Hydrodynamic evolution and characteristic scales}\label{sec:hydro}
We consider the interaction of SN ejecta with a dense stellar wind
that follows a standard profile $\rho_w = D r^{-2}$, where $D$ is a
constant and $r$ is the radius. This wind profile implies $\tau
\propto r^{-1}$, where $\tau$ is the optical depth of the wind at
the location of the shock. It also implies a photon diffusion time
scale to the observer, that is independent of the radius (up to a
logarithmic factor), $t_{diff}\sim \kappa D/c$, where $\kappa$ is
the cross section per unit of mass (throughout the paper we assume
that the gas which dominates the optical depth is fully ionized, and
take $\kappa=0.34 {\rm~cm^2 g^{-1}}$).

During the interaction, a forward shock is driven into the wind
while a reverse shock is driven into the ejecta.
\cite{Chevalier1982} derives a self similar solution of the
forward-reverse shock structure, assuming that both the ejecta and
the wind density profiles are power-laws:
\begin{equation}\label{eq:r(t)}
r(t)\propto t^{\frac{n-3}{n-2}}
\end{equation}
for $\rho_w \propto r^{-2}$ and $\rho_{ej} \propto v^{-n}$. The
ejecta density profile through which the reverse shock is
propagating is determined by the SN shock that crosses the stellar
envelope. The density profile of the fastest moving ejecta can be
approximated by a power-law with an index\footnote{The mass profile
for $m \ll M_{ej}$ is $m\propto
v^{-\frac{\tilde{n}+1}{0.19\tilde{n}}}$ where $\tilde{n}$ (not to be
confused with n) is the power-law index of the pre-explosion stellar
density profile near the edge. This profile translates, under free
expansion, to $\rho_{ej} \propto
t^{-3}v^{-3-\frac{\tilde{n}+1}{0.19\tilde{n}}}$, or $n\approx10$ for
a radiative envelope and $n\approx12$ for a convective one} $n
\approx 10-12$ for all progenitor types. The ejecta density profile
flattens towards the slower layers which contain most of the ejecta
mass and energy, \citep[e.g.][]{Matzner_McKee1999}. As the reverse
shock approaches these layers the evolution is no longer strictly
self-similar, but it can be approximated by the self-similar
solution using $n=7$
\citep[e.g.,][]{Chevalier_Irwin2011,Chevalier_Irwin2012}. Given that
during the reverse-forward shock interaction phase $n \gtrsim 7$ we
hereafter approximate $\frac{n-3}{n-2}\approx 1$ during this phase.
Finally, once the reverse shock ends crossing the ejecta only the
forward shock remains, entering a Sedov-Taylor phase if the shock is
adiabatic or a snowplow phase if it is radiative.

The density power-law indices of the ejecta and the wind are set by
the pre-explosion stellar evolution and are not expected to vary
much between various SN progenitors. The difference in the
hydrodynamical evolution during the forward-reverse shock phase
between various SNe is set by the two normalization factors of these
density profiles. These could be determined by two observables such
as the duration of the breakout pulse, $t_{bo}$, which is set by the
wind density only, $t_{bo}\sim\kappa D/c$
\citep{Ofek+2010,Chevalier_Irwin2011,Balberg_Loab2011}, and the
breakout luminosity $L_{bo}$. Below we present the evolution during
the interaction phase using $t_{bo}$ and the shock breakout velocity
$v_{bo}$, (instead of $L_{bo}$) which is related to the observables
via $v_{bo}=\left(\frac{\kappa}{4\pi
c}L_{bo}t_{bo}^{-1}\right)^{1/3}$ \citep{Balberg_Loab2011}. The
evolution after the interaction phase ends depends on three
parameters and we present it using $t_{bo}$, the total explosion
energy, $E$, and ejecta mass, $M_{ej}$.

The breakout from the wind takes place once $\tau_{bo}\approx
c/v_{bo}$, where $\tau_{bo}$ is the wind optical depth to the
observer at the breakout radius. This sets the mass swept by the
shock at the time of the breakout:
\begin{equation}\label{EQ m_bo}
m_{bo} \approx \frac{4\pi c}{\kappa}v_{bo}t_{bo}^2 \approx
5\times10^{-3}M_{\odot}v_{bo,9}t_{bo,d}^2 ~~,
\end{equation}
where $v_{bo,9}=\frac{v_{bo}}{10^9 {\rm~ cm/s}}$ and
$t_{bo,d}=\frac{t_{bo}}{1 {\rm~day}}$. During the ejecta-wind
interaction the wind mass collected by the forward shock is
comparable to the ejecta mass collected by the reverse shock.
Therefore, $v_{bo}$ can be found using the velocity profile of an
ejecta released by a SN explosion (e.g., from
\citealt{Nakar_Sari2010}; hereafter NS10), and requiring that a wind
mass collected by the forward shock until the breakout, $m_{bo}$, is
accelerated to $v_{bo}$ by the explosion. Equations A2 and A4 from
NS10 provide the ejecta velocity profile $v(m) \approx 3 \cdot 10^8
{\rm cm/s}
\left(\frac{m}{M_{ej}}\right)^{-\frac{0.19\tilde{n}}{\tilde{n}+1}}
\left(\frac{E_{51}}{M_{10}}\right)^{1/2}$, where
$M_{10}=\frac{M_{ej}}{10 M_\odot}$, $E_{51}=\frac{E}{10^{51}
{\rm~erg}}$ and $\tilde{n}$ (not to be confused with n) is the
power-law index of the pre-explosion stellar density profile near
the edge ($\tilde{n}=3$ for a fully radiative envelope and
$\tilde{n}=1.5$ for a fully convective one). Plugging Equation
\ref{EQ m_bo} into this velocity profile we find:
\begin{equation}\label{EQ v_bo}
    v_{bo} \approx 10^9 {\rm~cm/s~} M_{10}^{-0.35} E_{51}^{0.45}
    t_{bo,d}^{-0.2},
\end{equation}
where the dependence on $\tilde{n}$ is weak. This equation is
applicable until $m_{bo}=M_{ej}$. It implies that for typical SNe
(e.g., not extremely high or low $E_{51}/M_{10}$) and in the range
of $t_{bo}$ that we consider ($\gtrsim 1$ day) the breakout velocity
does not vary much. Thus, we consider here only a narrow range of
$v_{bo,9} \approx 0.3-1.5$. The upper limit of this range ensures
that shocks that follow breakouts with timescales of days or longer
remain fast cooling (i.e., the shock heated plasma radiate its
energy within a dynamical time) at all times (see section
\ref{sec:tslow}). When studying the post breakout evolution
we further limit the shock velocity by studying scenarios where the
unshocked wind electrons maintain thermal equilibrium with the
diffusing radiation, which typically implies $v_{bo,9} \lesssim 1$.
Note that since $v(m)$ is independent of the progenitor radius (the
radius determines only the maximal ejecta velocity), Equation
\ref{EQ v_bo} is independent of the radius. Since Equation \ref{EQ
v_bo} is also only weakly sensitive to the exact density profile
near the stellar edge, it is applicable to red and blue supergiant
as well as Wolf-Rayet progenitors. Thus, if the shock is breaking
out from a thick wind, the wind density is the parameter that
affects the observed radiation the most, and the progenitor type
cannot be identified based on the light curve alone.

An important time scale, which affects the hydrodynamic evolution,
is the time at which the wind mass collected by the forward shock,
$m$, is comparable to the total ejecta mass. At that point the
reverse shock stops playing a role and the forward shock is slowing
down rapidly. The deceleration rate depends on whether the shock is
adiabatic,  or a significant fraction of the post-shock internal
energy is radiated away. If it is adiabatic then the evolution
follows the Sedov-Taylor self similar solution. If, on the other
hand, all the internal energy behind the shock is radiated within a
dynamical time scale, then the shock enters the so called snowplow
phase. We consider only fast cooling shocks, which are in the
snowplow regime when $m \geq M_{ej}$ and $t
> t_{bo}$.

Next we find the time of transition to snowplow, $t_{SP}$, in case
that $t_{bo}<t_{SP}$. Equation \ref{eq:r(t)} for $r(t)$ can also be
applied for the snowplow phase using $n=4$ (the Sedov-Taylor phase
matches $n=5$)\footnote{Note that using $n=4$ in Equation
\ref{eq:r(t)} provides the correct evolution in the snowplow regime,
although \cite{Chevalier1982} derived it for the interaction phase
only.}. Therefore, in case that $t_{bo}<t_{SP}$ the swept mass after
breakout accumulates as
\begin{equation}
m(t)\approx \left\{\begin{array}{lr}
                     m_{bo} \frac{t}{t_{bo}} & t_{bo}<t<t_{SP}\\
                     M_{ej} \left(\frac{t}{t_{SP}}\right)^{1/2} &
                     t_{SP}<t
                   \end{array}\right.
\end{equation}
Since the snowplow regime begins once $m=M_{ej}$, it follows from
Equation \ref{EQ m_bo} that if $t_{bo}\lesssim 80
M_{10}^{0.75}E_{51}^{-0.25}$ d then $t_{bo}<t_{SP}$, where we
approximate the shock velocity at the transition to the snowplow
phase with $\sqrt{2E/M_{ej}}$. The transition time to snowplow phase
occurs then at
\begin{equation}\label{eq:t_SP}
t_{SP} = 7 \times 10^3 ~ t_{bo,d}^{-1}M_{10}^{3/2} E_{51}^{-1/2}
\,\rm{d}
\end{equation}
If $m=M_{ej}$ before breakout, then the evolution follows a
Sedov-Taylor expansion prior to breakout and assumes a snowplow
evolution afterwards. In such case we say that $t_{SP}<t_{bo}$ and
the mass behind the forward shock accumulates as
\begin{equation}
m(t)\approx m_{bo} \left(\frac{t}{t_{bo}}\right)^{1/2}~~~~~
t_{SP}<t_{bo}<t
\end{equation}

\section{Temperature and luminosity evolution}\label{sec:temperature}

\subsection{Bolometric luminosity}

The bolometric luminosity of a fast cooling shock is $L\approx 2\pi
r^2 \rho v^3\propto v^3\propto t^{-\frac{3}{n-2}}$. Therefore, if
the breakout takes place before $t_{SP}$ then
\begin{equation}\label{eq:L(t)}
L(t)\sim \left\{ \begin{array}{ll}
 5 \times 10^{43}\,t_{bo,d}\,v_{bo,9}^3 \left(\frac{t}{t_{bo}}\right)^{-0.3}\,\rm{\frac{erg}{s}} & t_{bo}<t\ll t_{SP}\\
 2 \times10^{42} \,t_{bo,d}M_{10}^{-3/2} E_{51}^{3/2}
\left(\frac{t}{t_{SP}}\right)^{-1.5}\,\rm{\frac{erg}{s}} &
t_{SP}\lesssim t
\end{array}\right.
\end{equation}
while if $t_{bo}>t_{SP}$ then
\begin{equation}\label{eq:Lt2}
L(t)\sim  5 \times 10^{43}\,t_{bo,d}\,v_{bo,9}^3
\left(\frac{t}{t_{bo}}\right)^{-1.5}\,\rm{\frac{erg}{s}} ~~;~~
 t_{SP}<t_{bo}<t .
\end{equation}
Note that once the radiation escape time becomes shorter than the
dynamical time (i.e., after breakout), then as long as the shock is
cooling fast, the bolometric luminosity is simply the kinetic energy
flux through the shock. In contrast to regular SNe, it is
independent of the gas opacity, and thus insensitive to processes
such as recombination or to the gas metallicity.

\subsection{Breakout temperature}\label{sec:breakout}

The first signal to be observed is the breakout pulse, releasing
over a diffusion time $t_{diff}\sim t_{bo}$ the energy accumulated
during the interaction with the wind prior to the breakout. The
pre-breakout forward shock is radiation mediated, but the photons
that mediate the shock and dominate the energy density behind the
shock cannot escape before breakout. As the breakout occurs, these
photons start leaking towards the observer and their observed
temperature is set by their interaction with the unshocked wind
through which they diffuse.

The unshocked wind is heated by the diffusing radiation. It produces
photons, mostly by free-free, which share energy with the diffusing
radiation. The breakout temperature is therefore set by the ability
of the unshocked wind to produce photons over the radiation
diffusion time. The radiation energy density at any point along the
diffusion path is $\epsilon_{rad}\sim\frac{L \tau}{cr^2}$. We define
$T_{BB} \equiv (\epsilon_{rad}/a)^{1/4}$ , where $a$ is the
radiation constant. If photons are abundant enough behind the shock
and the radiation is in thermal equilibrium, then the photon
temperature at the shock is $T_{BB}$. The observer is seeing,
however, a lower temperature. The reason is that thermal equilibrium
is also kept at radii that are larger than the shock radius and
$\tau$ is lower, where the radiation energy density is lower. Thus,
the observed temperature is set by the radiation energy density at
the largest radius at which thermal equilibrium is kept. If, in
contrast, the photons at the shock are out of thermal equilibrium,
then the observed temperature can be much higher than $T_{BB}$, and
is set by the number of photons that are generated over the
diffusion time. This process is the same one as in the case of a
shock breakout from a stellar surface where no wind is present, and
is discussed in length in \cite{Katz+2010} and NS10.

Assuming that photons and electrons in the unshocked wind have the
same temperature, the observed temperature can be determined using
the coupling coefficient $\eta \equiv 5 \times 10^{-36}
T_{BB}^{3.5}\rho^{-2} t_{diff}^{-1}$ defined in NS10, where all
quantities are in c.g.s. units. Here $\eta$ is the ratio of the
photon density needed to maintain thermal equilibrium to the photons
produced per unit of volume during the diffusion time, assuming that
the electrons' temperature is $T_{BB}$. This value of $\eta$ assumes
that free-free emission dominates photon production. If bound-free
opacity for $T_{BB}$ photons, $\kappa_{bf}$, is higher than
free-free opacity, $\kappa_{ff}$, then $\eta$ is lower by the factor
$\kappa_{bf}/\kappa_{ff}$ (see discussion in NS10). When the value
of $\eta$ at the shock radius is $\eta_s < 1$, the radiation is in
thermal equilibrium and the temperature is set by the electrons at
the radius where\footnote{We assume here that the Thompson cross
section is larger than the free-free absorption cross section and
that $\tau>1$. This is true during the breakout and later while the
soft component is dominant} $\eta=1$.

\begin{figure}[t!]
\includegraphics[width=9cm]{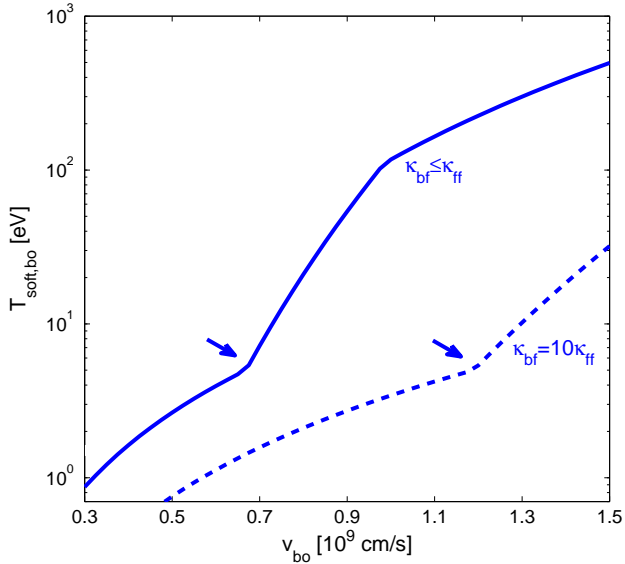}
\caption{The observed breakout temperature as a function of the
breakout velocity for $t_{bo}=20$ d, demonstrating the strong effect
of deviation from thermal equilibrium on the observed temperature.
Departure from equilibrium is indicated by arrows. The solid
line assumes photons are produced solely by free-free emission (low
metallicity). The relatively small range of relevant velocities
$0.3<v_{bo,9}<1.5$ corresponds to a range spanning over 2.5 orders
of magnitude in temperature, $10^4 K<T_{bo}<5 \times 10^6 K$.
The dashed line assumes a bound-free opacity ten times that
of free-free, resulting in departure from thermal equilibrium at
higher velocity and lower breakout temperatures. The latter is
because thermal equilibrium is kept farther ahead of the shock,
thereby reducing the observed temperature, although the shock
temperature is independent of the opacity.}\label{fig:Tbo}
\end{figure}

At breakout, $aT_{BB}^4/3=(6/7)\rho_{w,bo}v_{bo}^2$, implying
\begin{equation}
T_{BB, bo}=\left(\frac{18c}{7\kappa at_{bo}}\right)^{1/4}\approx
10^5 t_{bo,d}^{-1/4}\,\rm{K} ,
\end{equation}
and yielding\footnote{Note that calculating $\eta$ for the
configuration of a wind breakout is slightly different than in the
steady state planar case discussed by \cite{Weaver1976} and
\cite{Katz+2010}. The reason is that in a steady state shock photons
that are diffusing from the downstream back to the shock, over the
shock crossing time, are produced by a mass that is larger by the
shock compression ratio, 7 in this case, than the mass in the shock
transition layer. In the case of a wind breakout the mass behind the
shock is comparable to the mass in the shock transition layer at the
time of breakout. Therefore $\eta_{bo}$ is larger by a factor of
$\approx$7 compared to the one that is calculated for the same shock
velocity in steady state. As a result the radiation behind the wind
breakout shock falls out of thermal equilibrium already when the
shock exceeds 7000 $\rm{km\,s^{-1}}$ (for $\kappa_{bf} <
\kappa_{ff}$) instead of 15,000 $\rm{km\,s^{-1}}$ in the steady
state case.}
\begin{equation}\label{eq:eta}
\eta_{bo} \approx 5
\left(1+\frac{\kappa_{bf}}{\kappa_{ff}}\right)^{-1}
v_{bo,9}^4 t_{bo,d}^{1/8},
\end{equation}
where free-free is assumed to be the main source of photons, (and
absorption opacity). At any given time $\eta \propto r^{11/8}$,
implying that if $\eta_{bo}\ll 1$, then the observed temperature of
the breakout pulse is $T_{BB,bo} \eta_{bo}^{6/11}$, yielding an
observed breakout temperature
\begin{equation}\label{eq:TsoftBO}
T_{bo,obs}(\eta_{bo}<1)\approx 2 \times 10^4\,
\left(1+\frac{\kappa_{bf}}{\kappa_{ff}}\right)^{-\frac{6}{11}}
\left(\frac{v_{bo}}{3000\,\rm{\frac{km}{s}}}\right)^{2.2}
t_{bo,d}^{-0.2} \,\rm{K}.
\end{equation}
In contrast, when $\eta_{bo}\gg 1$, then the observed breakout
temperature is much higher than $T_{BB}$ and is given by
$T_{BB,bo}\eta_{bo}^2\xi^{-2}$ (Equation 13 in NS10). This equation
is implicit since the factor $\xi$ depends on $T_{BB,bo}$. $\xi$ is
a logarithmic factor which accounts for the contribution of inverse
Compton in bringing the temperature closer to thermal equilibrium
($\xi$ always $\geq 1$). It is discussed in length in NS10 and its
value is $\sim 1$ at low temperatures ($\lesssim 10^6$ K) and $\sim
10$ at high temperatures ($\gtrsim 10^7$ K). In the scenario that we
discuss here the calculation of $\xi$ is slightly modified compared
to the one described in NS10, where the softest photon that is
thermally coupled to the electrons is set by the requirement that
its energy is doubled before it is absorbed by free-free (see their
Equation 11). Here, where the optical depth of the wind can be low
enough, the limiting factor may become instead the low Compton
y-parameter of the wind. In our calculations below we use both
requirements (high enough y-parameter and no free-free absorption)
to find the softest photons that are coupled by IC to the electrons.

Figure \ref{fig:Tbo} depicts the observed breakout temperature as a
function of the breakout velocity for $t_{bo}=20$ d. The solid line
describes a wind composition such that photons production (and
absorption) is dominated by free-free emission. In that case the
critical breakout velocity, above which thermal equilibrium cannot
be maintained, is $\approx 7000$ $\rm{km\,s^{-1}}$. It is evident
that $T_{bo,obs}$ depends strongly on deviation from thermal
equilibrium, rising by more than an order of magnitude when $v_{bo}$
is increased by a factor of 2. Thus, for $v_{bo,9}=0.3$ we find
$T_{bo,obs} \approx 10^4$ K while if $v_{bo,9}=1.5$ then $T_{bo,obs}
\approx 5 \times 10^6$ K. The effect of high metallicity is
demonstrated by the dashed line in Figure \ref{fig:Tbo}, describing
the observed breakout temperatures in case that
$\kappa_{bf}/\kappa_{ff}=10$ (and $\eta_{bo}$ is smaller by that
factor). Evidently, a higher metallicity implies a thermal
equilibrium also at higher velocities. Note, that $T_{bo,obs}$ at
which thermal equilibrium breaks is almost independent of the exact
opacity ($5 \times 10^4$ K in Figure \ref{fig:Tbo}). It also depends
weakly on $t_{bo}$ and is always $<10^5$ K. Since the velocity range
that we consider here is exactly at the point where thermal
equilibrium is marginal, we expect some of the breakouts in this
range to be in equilibrium with $T_{bo,obs} \sim 10^4-10^5$ K,
therefore bright in optical/UV, and some to be out of thermal
equilibrium with $T_{bo,obs}> 10^6$ K, bright in soft X-rays, but
faint in optical.

\subsection{Post breakout evolution}\label{sec:post}

Following breakout, the collisionless shock that develops heats
protons to high temperatures. The protons then transfer heat to
electrons, which are cooled by radiation. The balance between the
electrons heating and cooling rates sets the electron temperature,
 during the post-shock cooling process \citep{Katz+2011,Murase+2011}. The
breakout balance temperature is\footnote{This equation assumes that
the shock does not couple the electrons to the protons and that the
electron heating is by Coulomb interactions. In this case the
electrons Coulomb heating and radiative cooling are both
proportional to protons kinetic energy and the resultant electrons
balance temperature is independent of the shock velocity. If instead
the shock couples the electrons to the protons then electrons are
heated by the shock to $T_p$ and then cool down until the radiative
cooling and Coulomb heating is balanced. Since the difference
between $T_p$ and $T_e$ is not large under any of the two scenarios
their predictions for the observed signature are similar.} $\sim 60$
keV, independent of the shock velocity. If the protons temperature,
$k_BT_p\approx\frac{3}{16}m_pv^2\approx200v_9^2$ keV where $m_p$ is
the proton mass, is lower than the implied balance temperature, then
$T_e\sim T_p$. Thus
\begin{equation}
k_BT_{e,bo} \sim \min\left[200v_{bo,9}^2,60\right]\,\rm{keV} .
\end{equation}
It is straight forward to show that after breakout $T_e$ approaches
$T_p$ (if they where not similar already at breakout). Now, since we
consider $v_{bo,9}=0.3-1.5$, the electron and proton temperatures
are not very different already during the breakout, and at
$t<t_{SP}$ the proton and electron temperatures are approximately
constant. At $t>t_{SP}$ the proton temperature drops, and $T_e$
follows it. Therefore, the electron temperature can be approximated
as
\begin{equation}\label{eq:Te}
k_BT_e(t) \sim \left\{\begin{array}{lr}
                 60~{\rm keV} & t<t_{SP} \\
                 20E_{51} M_{10}^{-1} \left(\frac{t}{t_{SP}}\right)^{-1}~{\rm keV}  &~~ t>t_{SP}
               \end{array}\right. .
\end{equation}

While the electrons behind the collisionless shock are hot,
the observed radiation at the breakout, and soon after it, is much
softer \citep{Chevalier_Irwin2012}. The reason is that energy of the
shocked electrons is deposited into photons that diffuse through the
optically thick colder gas that lies ahead of the shock. During the
diffusion, the temperature of hard photons is reduced via Compton
scattering on the unshocked, lower temperature, wind electrons.
Moreover, the outgoing radiation heats the unshocked gas along the
diffusion path, which in turn produces photons via free-free and
bound-free emission. In the case that we consider here the gas ahead
of the shock is in thermal equilibrium and its temperature is in the
optical-UV range. The generated optical-UV photons dominate the
radiation energy density and are the main cooling source, via
inverse Compton (IC) of the hot shocked electrons. Thus, the energy
in hard photons, which are emitted by free-free emission of the hot
shocked electrons, is suppressed farther by the free-free to IC
cooling rate ratio.

Hence, if the electrons of the unshocked wind maintain
thermal equilibrium with the diffusing radiation (not to be confused
with the hot post-shock electrons that are always out of thermal
equilibrium) the observed spectrum which follows the breakout pulse
will have two distinct components, a soft one and a hard one.
Photons of the softer component, which dominates the luminosity at
early times, are produced by the unshocked electrons and their main
energy source is IC over the hot shocked electrons. The hot
electrons themselves radiate hard free-free photons at $T_e$. The
observed temperature of these hard photons is determined by the
energy loss of a single photon while it diffuses out of the wind and
their fraction of the total luminosity is set by the product of this
energy loss and the free-free to IC cooling rate ratio. Below we
describe the temperature evolution of the soft component,
$T_{soft}$, and the temperature and luminosity of the hard
component, $T_{\rm{hard}}$ and $L_{hard}$, while $T_{soft}$ is in
thermal equilibrium.

\subsubsection{Soft component}\label{sec:soft}

The shock-heated electrons and protons are confined to a narrow
layer behind the shock. Although they are the source of the observed
luminosity, they do not necessarily set the typical photon
temperature. This temperature is set during the diffusion of the
photons towards the observer through the unshocked wind, as long as
it is optically thick\footnote{Fast cooling dictates that the
shocked plasma is cooling, forming a dense cold layer behind the hot
shock heated plasma. This cold layer does not affect the observed
soft component. The reason is that after the breakout the photons
escape on a time that is shorter than the dynamical time, and
therefore the energy density behind the shock is not dominated by
radiation. Since the shock is radiative the shocked plasma is
compressed to keep the pressure balance and once free-free cooling
becomes dominant, its cooling rate increases with the compression
and a run-away cooling decouples the plasma temperature from the
rest of the system, preventing it from contributing a significant
number of photons to the radiation field that cools the shock heated
plasma.}.

The electrons of the unshocked wind are much colder than
the $\sim 60$ keV electrons immediately behind the shock. These
unshocked electrons are heated by the diffusing radiation, and they
produce photons which share energy with the diffusing radiation.
During the early evolution after the breakout, the major part of the
$\sim 60$ keV electrons layer cooling occurs as photons produced by
the unshocked electrons diffuse back into the hot electrons layer
behind the shock, and heat up by inverse Compton over this hot
layer. These upscattered photons also serve during this stage as the
main heating source of the unshocked electrons, through absorption
of photons with similar temperature, and possibly also through
Compton collisions with photons upscattered to $m_ec^2/\tau^2$,
where Compton losses are substantial. The balance between the
heating and the cooling of the unshocked electrons determines their
temperature.

If unshocked electrons are able to produce a sufficient amount of
photons during the available diffusion time, then the diffusing
radiation is in thermal equilibrium. The condition for thermal
equilibrium during the post-breakout evolution is similar to that
during the breakout, namely $\eta<1$. Thermal equilibrium also
implies that each photon with $T \sim T_{BB}$ is absorbed at least
once during its diffusion through the unshocked electrons. This
ensures that the unshocked electrons and the diffusing photons share
a single temperature and the observed temperature is simply $T_{BB}$
at the outermost layer that can maintain thermal equilibrium, namely
the layer where $\eta=1$. At breakout, the blackbody temperature of
the diffusing radiation just ahead of the shock is
$T_{BB,s}=T_{BB,bo}$, and the thermal coupling coefficient of the
electrons at this location is $\eta_s=\eta_{bo}$. Following the
breakout $T_{BB}$ and $\eta_s$ (both measured at the shock radius)
evolve as
\begin{equation}
T_{BB,s}(t)\propto\left(\frac{L\tau}{r^2}\right)^{1/4}\propto
t^{-3/4} ,
\end{equation}
independent of $n$, and
\begin{equation}
\eta_s(t)\propto\frac{T_{BB,s}^{3.5}}{\rho^2}\propto T_{BB}^{3.5}r^4
\propto\left\{ \begin{array}{ll}
t^{1} & n=12\\
t^{0.6} & n=7\\
t^{-0.6} & n=4
\end{array}\right. .
\end{equation}
The evolution of the observed soft component temperature ($T_{BB}$
where $\eta=1$) while thermal equilibrium is maintained is
therefore:
\begin{equation}
T_{soft}(t, \eta_s<1)\propto T_{BB,s} \eta_s^{6/11} \propto \left\{
\begin{array}{ll}
t^{-0.2} & n=12\\
t^{-0.45} & n=7\\
t^{-1.1} & n=4
\end{array}\right. .
\end{equation}
Thus, while $\eta_s<1$, the observed temperature decreases slowly at
$t<t_{SP}$ and it drops linearly with $t$ during the snowplow phase
for as long as $k_BT_{soft}\gtrsim 6,000$ K, beyond which
recombination takes place and the temperature drop stops.

The main difference between the scenario discussed above
($\eta_s<1$), and the one in which the radiation ahead of the shock
deviates from thermal equilibrium, i.e. $\eta_s>1$, is that
absorption of the photons that carry most of the luminosity by
unshocked electrons stops playing a role. This has two important
implications. First, the energy of a single photon can significantly
grow by IC over the shocked hot electrons without being absorbed.
The limit on the observed energy of such IC upscattered photons is
set by three factors: Compton losses at $m_ec^2/\tau^2$, escape time
from the system and photoabsorption of extreme UV/soft X-rays by
partially ionized pre-shock metals. Second, the coupling between the
diffusing photons and unshocked electrons is done  by Compton
scattering only. This implies, in the parameter space that we
consider, that the unshocked electrons and diffusing photons cannot
maintain a single, well defined temperature. As a result the photons
produce a non trivial spectrum between the temperature of the
unshocked electrons and $m_ec^2/\tau^2$ and there may be no clear
distinction between soft and hard components. Calculating the exact
observed photon spectrum in that case is beyond the scope of this
paper.

In general, our analysis of the soft component is not
applicable anymore when the first of the following three takes
place: (i) recombination becomes important, at which point $T_{soft}
\approx 6000$ K (ii) the hard component carries most of the
luminosity (which always precedes the inapplicability of the
diffusion approximation due to low wind optical depth) (iii)
$\eta_s>1$.

\subsubsection{Hard component}\label{sec:hard}
The soft component is accompanied by harder photons, which
can be generated by the hot electrons behind the shock in two ways,
(i) free-free emission (ii) repeated IC upscatter of photons from
the soft component. The total energy in IC upscattered photons is
exponentially suppressed when $\eta_s<1$, since only photons at the
exponential tail of $T_{soft}$ can be upscattered significantly
before being absorbed by electrons with $T_{soft}$. It may be
suppressed farther at $T \gg T_{soft}$ by photoabsorption of
partially ionized metals ahead of the shock. We therefore consider
here only hot electrons' free-free emission. At breakout, the main
cooling source of the hot electrons is IC of soft photons, yielding
a free-free to IC emissivity ratio
\citep[c.f.][]{Chevalier_Irwin2012}:
\begin{equation}\label{eq:ff_to_IC_bo}
\frac{\varepsilon^{ff,bo}}{\varepsilon^{IC,bo}}=\frac{\alpha
n_e\left(m_ec^2\right)^{3/2}}{4\sqrt{k_B
T_e}\epsilon_{rad}}\approx10^{-2}v_{bo,9}^{-2} ,
\end{equation}
where $\epsilon_{rad} \approx \rho v^2/2$ at breakout. After the
breakout, as long as $\tau>1$,
\begin{equation}\label{eq:ff_to_IC}
\frac{\varepsilon^{ff}}{\varepsilon^{IC}}(t)\propto\frac{1}{
T^{1/2}v^3\,\tau } \propto \left\{\begin{array}{ll}
t &  ~t<t_{SP}\\
t^{2.5} & ~t>t_{SP}\\
\end{array}\right. .
\end{equation}
Therefore, IC remains the dominant cooling source for a long time
after the breakout, suppressing the emission of hard photons. But
even the hard photons that are emitted at $T_e$ are not seen
directly by the observer as long as the wind optical depth is
significant. Compton losses during their diffusion towards the
observer limit their energy to $m_e c^2/\tau^2$ and farther reduces
the luminosity of the hard component by a factor $\tau^2 T_e/m_e
c^2$ \citep{Chevalier_Irwin2012}. Therefore, the observed
temperature of the hard photons is
\begin{equation}\label{eq:Thard}
\begin{array}{ll}
T_{\rm{hard}}(t)=\min\left[\frac{m_ec^2}{k_B\tau^2},T_e(t)\right]~.
\end{array}
\end{equation}

The luminosity of hard photons, counting only contribution
of free-free emission of the hot shocked electrons, is:
\begin{equation}
L_{\rm{hard}}(t)\sim
L(t)\frac{T_{\rm{hard}}(t)}{T_e(t)}\min\left[1,\frac{\varepsilon^{ff}}{\varepsilon^{IC}}(t)\right]
. \label{eq:L_hard}
\end{equation}
Equations \ref{eq:Te}, \ref{eq:ff_to_IC_bo}, \ref{eq:Thard}  and
$\tau_{bo}=c/v_{bo}$ imply that when the breakout is at thermal
equilibrium, this luminosity is always
\begin{equation}
L_{\rm{hard},bo} \sim 10^{-4} L_{bo}.
\end{equation}
Equations \ref{eq:L(t)} and \ref{eq:ff_to_IC} implies that at
$t<t_{SP}$, while $L_{\rm{hard}} \ll L$, it first evolves rapidly
and then slows down, within the range :
\begin{equation}\label{LhardEarly}
L_{\rm{hard}}(t<t_{SP}) \propto t^{2.5}\rightarrow t .
\end{equation}
If the hard component is still subdominant during the transition to
the snowplow phase, or if the breakout takes place during this
phase, then its fraction out of the total luminosity rises quickly
and multiplying this fraction with  $L(t)$:
\begin{equation}\label{LhardLate}
L_{\rm{hard}}(t>t_{SP}) \propto t^{3}\rightarrow t^{0.5},
\end{equation}
until it becomes comparable to the bolometric luminosity.

Finally, Equation \ref{eq:L_hard} shows that right after
breakout the hard photons from the hot electron's free-free emission
carries only a tiny fraction ($\sim 10^{-4}$) of the bolometric
luminosity (which is seen in optical /UV). It implies, that even
though the contribution of repeated IC scattering of $T_{soft}$
photons is strongly suppressed, it may still dominate the luminosity
of the hard component. Therefore, while $\eta_s \ll 1$ implies,
$L_{\rm{hard},bo} \ll L_{bo}$, the fraction may be higher than
$10^{-4}$ right after the breakout.

\subsection{Criterion for  fast cooling}\label{sec:tslow}
We assumed above that the shocked plasma is cooling fast (within a
dynamical time). Here we find when this assumption holds. A
sufficient requirement for fast cooling is that free-free emission
cools the hot electrons over a dynamical time. In case the Compton
y-parameter drops below unity, it is also a necessary condition. The
free-free cooling time at breakout is much shorter than the
dynamical time for our considered range of $v_{bo}$:
\begin{equation}\label{eq:tslow_bo}
\frac{t_{cool,bo}^{ff}}{t_{bo}}\approx
\frac{1}{25}v_{bo,9}^4T_{e,60}^{-1/2}
\end{equation}
Past breakout, it evolves as
\begin{equation}\label{eq:tslow}
\frac{t_{cool}^{ff}}{t}\propto
v^2n_e^{-1}T_e^{-1/2}t^{-1}\propto\left\{
\begin{array}{ll}
t^{0.6} & n=12\\
t^{0.2} & n=7\\
t^{-0.5} & n=4
\end{array}\right.
\end{equation}
Thus, if the shock is cooling fast at $t_{SP}$ then it continues to
cool fast also later. Requiring $t_{cool}^{ff}(t_{SP})/t_{SP}<1$ and
taking $t_{cool}^{ff}/t \propto t^{0.6}$  ($n=12$) all the way from
breakout to $t_{SP}$ shows that the shock enters the snowplow phase
cooling fast as long as $t_{bo} \gtrsim 6 v_{bo,9}^{10/3}
M_{10}^{3/4} E_{51}^{-1/4}\,\rm{d}$. This calculation over estimates
the minimal $t_{bo}$  in equation \ref{eq:tslow} by a factor of
$\sim 3$ since $n$ varies from $n \lesssim 12$ at breakout to $n
\approx 4$ at $t_{SP}$. Therefore, in the range of breakout
velocities that we consider, the shock is always fast cooling for
breakout times that are days or longer.

\section{Light curves of the different regimes}
The light curve and spectral evolution depend on three time scales.
The first two are $t_{bo}$ and $t_{SP}$, which affect the
hydrodynamics. The third timescale, which we denote $t_{\rm{hard}}$,
marks the time at which the hard component becomes the dominant one,
thus affecting the spectral evolution. For breakout times of a day
or longer, the order of appearance of these three characteristic
timescales depends primarily on the breakout time $t_{bo}$.

Since at breakout the hard component is always a small fraction of
the total luminosity, $t_{bo}<t_{\rm{hard}}$, the evolution depends
on the location of $t_{SP}$ with respect to $t_{bo}$ and
$t_{\rm{hard}}$. One regime is $t_{SP}<t_{bo}<t_{\rm{hard}}$. The
criterion for this regime was found in section \ref{sec:hydro}. When
$t_{bo}<t_{SP}$, there are two additional regimes, separated by the
limiting case of $t_{SP}=t_{\rm{hard}}$. In order to find the
separating criterion we use Equation \ref{eq:L_hard}. $t_{hard}$ is
the earliest time that satisfies $L_{hard} \approx L$, namely that
both $\tau^2 \approx m_ec^2/k_BT_e$ and $\varepsilon^{ff} \approx
\varepsilon^{IC}$ are satisfied. Setting $t_{SP}=t_{\rm{hard}}$ the
condition $\tau^2 \approx m_ec^2/k_BT_e$, which is the last to be
satisfied for typical parameters, implies $\tau \approx
5E_{51}^{1/2}M_{10}^{-1/2}$, where we used
 $v(t_{SP}) \approx \sqrt{2E/M_{ej}}$ and Equation
\ref{eq:Te}. Using Equation \ref{eq:t_SP} and setting
$\tau=5E_{51}^{1/2}M_{10}^{-1/2}$ one finds that
$t_{SP}=t_{\rm{hard}}$ when $t_{bo,d} \approx
20M_{10}^{0.75}E_{51}^{-0.25}$.  We thus obtain the following three
regimes with respect to $t_{SP}$:
\small
\begin{equation}
\begin{array}{lc}
t_{SP}<t_{bo} & t_{bo,d}\gtrsim 80M_{10}^{0.75}E_{51}^{-0.25}\\
&\\
t_{bo}<t_{SP}<t_{\rm{hard}} & ~~~~20M_{10}^{0.75}E_{51}^{-0.25}\lesssim t_{bo,d}\lesssim 80M_{10}^{0.75}E_{51}^{-0.25}\\
&\\ t_{bo}<t_{\rm{hard}}<t_{SP} & 1\lesssim t_{bo,d}<
20M_{10}^{0.75}E_{51}^{-0.25}
\end{array}
\end{equation}
\normalsize These regimes roughly match initial (breakout) $n$
values of 4, 7 and 12, respectively. Below we discuss the luminosity
and spectral evolution in each of these regimes.

\subsection{Late breakout ($t_{SP}<t_{bo}$)}
\begin{center}  $t_{bo,d}\gtrsim 80M_{10}^{0.75}E_{51}^{-0.25}$\end{center}

\begin{figure}[t!]
\includegraphics[width=8.5cm]{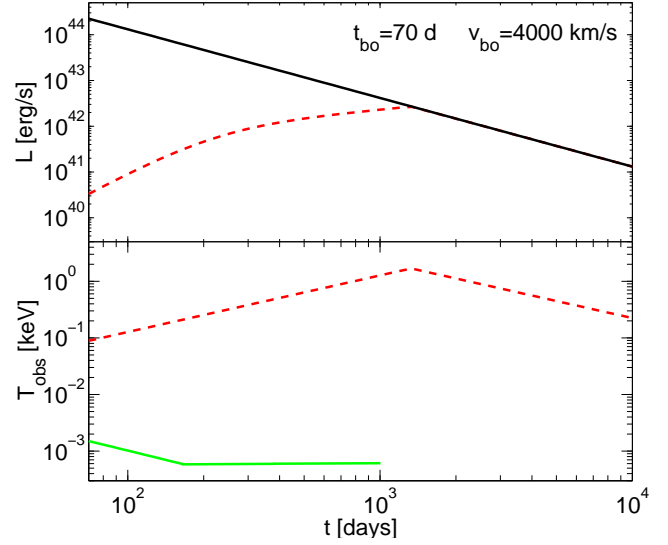}
\caption{A breakout within the snowplow regime, with $t_{bo}=70$ d
and $v_{9,bo}=0.4$ (matching $E_{51}=1.65$ and $m_{bo}=10M_\odot$),
showing a very bright flash in the optical-UV, with a comparable
rise time (not seen in the figure) and fall time. The temperature of
the soft component (calculated for $\kappa_{bf}\leq \kappa_{ff}$)
and the bolometric luminosity  ({\it solid lines}) decline as
$t^{-1.1}$ and $t^{-1.5}$ respectively. Temperature decline stops
around 6000 K, where recombination starts playing an important role.
The luminosity of hard photons that are generated by free-free
emission of the hot shocked electron and their temperature ({\it
dashed red lines}) are strongly suppressed around the breakout (see
text). $\geq$ keV photons are observed only more than a year after
the explosion. The breakout time and velocity are chosen to fit the
observation of SN 2006gy.}\label{fig:LateBO}
\end{figure}

If the progenitor went through an extreme mass loss episode just
prior to the explosion, the mass in the wind can be larger than the
ejecta mass. In such a scenario the breakout may take place weeks to
months after the explosion, near the time that, or even after, the
reverse shock is exhausted. The entire explosion energy is released
then around $t_{bo}$.  If the ejecta is several solar masses with
$\sim 10^{51}$ erg, the breakout velocity is $3000-5000$
$\rm{km\,s^{-1}}$. At these velocities $\eta_s<1$ throughout the
evolution even for $\kappa_{bf} \leq \kappa_{ff}$, implying a very
bright optical-UV flash, with a comparable rise time and fall time,
and a typical temperature of $\sim 1-5 \times 10^4$ K. Following
breakout, the luminosity and temperature of the soft component
decline as $t^{-1.5}$ and $t^{-1.1}$ respectively. The temperature
decline stops around 6000 K, where recombination starts playing an
important role. The breakout free-free hard component is strongly
suppressed, ($\sim 10^{-4}$ of the total luminosity) by the IC
cooling dominance and the high wind optical depth.
Contribution of IC photons to the hard component may
dominate over free-free at early time, but since $\eta_s \ll 1$
their luminosity will still be much lower than $L_{bol}$. Moreover,
the temperature of hard photons is limited during the breakout to
$\lesssim 0.1$ keV by Compton loses during the diffusion through the
unshocked wind \citep{Chevalier_Irwin2012}. The hardness and
luminosity of the hard component rise following the breakout, but a
significant emission above 1 keV starts only at $\sim 10t_{bo}$.
Luminous hard X-ray emission is expected in this regime only if the
breakout velocity is high, which requires either a low ejecta mass
$\sim M_\odot$ or a very high energy explosion $\sim 10^{52}$ erg.
Figure \ref{fig:LateBO} depicts the luminosity and temperature
evolution of a breakout near the snowplow phase with $t_{bo}=70$ d
and $v_{9,bo}=0.4$ where the evolution discussed above is seen. The
breakout time scale and velocity fit the observations of SN 2006gy
\citep{Ofek07,Smith07}, which was suggested by
\cite{Chevalier_Irwin2011} to be a shock breakout through a wind.
The luminosity and temperature  of the dominating soft component of
SN 2006gy are recovered with these $t_{bo}$ and $v_{bo}$. Similarly
to the conclusion of \cite{Chevalier_Irwin2012}, we find a strongly
suppressed X-ray luminosity, with keV photons appearing only years
after the explosion (assuming that the shock does not encounter the
edge of the wind at an earlier time).

Note that already during the breakout the mass of the wind is
comparable to the mass of the ejecta, and the mass accumulated by
the shock grows as $t^{1/2}$. Therefore, if the wind mass is not
much larger than the ejecta mass, the shock encounters the end of
the massive wind not too long after $t_{bo}$.

\subsection{Early breakout,  ($t_{bo}<t_{\rm{hard}}<t_{SP}$)}
\begin{center}$1\lesssim t_{bo,d}\lesssim 20M_{10}^{0.75}E_{51}^{-0.25}$\end{center}

\begin{figure}[t!]
\includegraphics[width=8.5cm]{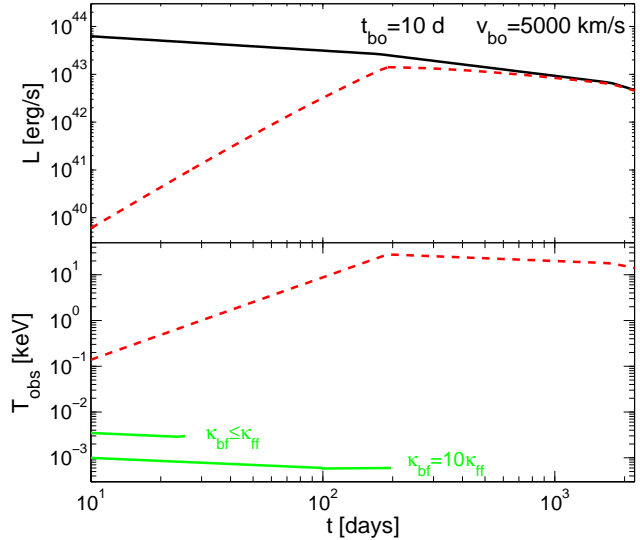}
\caption{Early breakout of $t_{bo}=10$ d and $v_{bo}=5,000$
$\rm{km\,s^{-1}}$, where $t_{SP}$ is calculated for $E_{51}=1$ and
$M_{10}=1.25$. The Bolometric luminosity and soft component
temperature are marked with {\it solid lines}. The temporal decay of
the bolometric luminosity is shifted from $t^{-0.3}$ (n=12) to
$t^{-0.6}$ (n=7) at $0.1t_{SP}$, to account for the flattening of
the ejecta profile once $m$ approaches $M_{ej}$. Such a breakout may
explain SNe like PTF 09uj \citep{Ofek+2010}. $T_{soft}$ is
plotted for two different absorption opacities. When
$\kappa_{bf}\leq\kappa_{ff}$ (low metallicity) then $\eta_{bo}=0.3$
and $T_{soft}$ decreases slowly before thermal equilibrium is broken
at day 25 (where we stop following it). In case that
$\kappa_{bf}=10\kappa_{ff}$ (higher metallicity) then
$\eta_{bo}=0.03$ and $T_{soft}$ remains in thermal equilibrium until
the soft component becomes negligible. The luminosity and
temperature of the hard photons are marked with {\it dashed lines}.
The luminosity includes only contribution from hot shocked
electrons' free-free emission, which is very faint at first, and
becomes brighter and harder quickly to dominate the flux from day
200 and on. Contribution from IC photons (not included in the hard
component here) may be important at early time, especially in case
that $\kappa_{bf}\leq\kappa_{ff}$ where thermal equilibrium is
marginal. This may result in significant X-ray emission also earlier
than day 200. Such events make excellent candidates for X-ray
searches $\sim$100-300 d after the SN explosion and possibly even
earlier. }\label{fig:EarlyBO}
\end{figure}

If breakout occurs early the shock velocity is larger, ranging from
5,000-10,000 $\rm{km\,s^{-1}}$ for typical SNe. For low metallicity,
i.e. $\kappa_{bf} \leq \kappa_{ff}$, the breakout emission is at a
marginal thermal equilibrium and a breakout temperature in the range
$10^4-10^6$ K is expected for SNe with typical energy and ejecta
mass. Figure \ref{fig:EarlyBO} depicts the luminosity and
temperature of the soft and hard components in a breakout with
$t_{bo}=10$ d and $v_{bo}=5,000$ $\rm{km\,s^{-1}}$. If metallicity
is low, the soft component is in marginal thermal equilibrium during
breakout, $\eta_{bo}=0.3$, resulting in $T_{bo,obs} \approx 25,000$
K. Following the breakout $\eta_s$ increases roughly linearly with
time, implying that it becomes harder for the radiation to maintain
thermal equilibrium. As a result $T_{soft}$ declines very slowly
until day 25, when thermal equilibrium cannot be maintained anymore.
If $\kappa_{bf}=10\kappa_{ff}$, then $\eta_{bo}=0.03$ and $T_{soft}$
remains in thermal equilibrium until the soft component becomes
negligible. In that case $T_{bo,obs} \approx 10,000$ K, and
$T_{soft}$ decreases slowly after breakout towards recombination
temperature. Hard photons from free-free emission carry a very small
fraction of the bolometric luminosity at first (only $\sim
10^{-4}$), and thus hard IC photons may dominae the hard component
at early times (especially when $\kappa_{bf} \leq \kappa_{ff}$ and
the thermal coupling is weak). The temperature of hard photons is
limited after breakout to $\sim 0.5$ keV. The hard component becomes
brighter and harder quickly, with hard X-rays and soft gamma-rays
dominating the flux after 200 days (and possibly earlier if
IC produces hard photons efficiently when $\eta_s>1$). Since the
luminosity decays slowly before $t_{SP}$, these events are good
candidates for X-ray searches $\sim$100-300 day after the SN
explosion, when $>$keV photons must dominate the luminosity and
before the snowplow phase starts. Finally, the luminosity increases
with breakout time. Therefore, breakouts with larger $t_{bo}$ (which
are still in this regime) produce brighter X-ray events.

Early breakout may explain SN PTF 09uj, as suggested by
\cite{Ofek+2010}. The breakout temperature and luminosity of the
case depicted in figure \ref{fig:EarlyBO} are consistent with the
observations \citep{Ofek+2010}.

\subsection{Intermediate breakout time, ($t_{bo}<t_{SP}<t_{\rm{hard}}$)}
\begin{center}$20M_{10}^{0.75}E_{51}^{-0.25}\lesssim t_{bo,d}\lesssim
80M_{10}^{0.75}E_{51}^{-0.25}$\end{center}

\begin{figure}[t!]
\includegraphics[width=8.5cm]{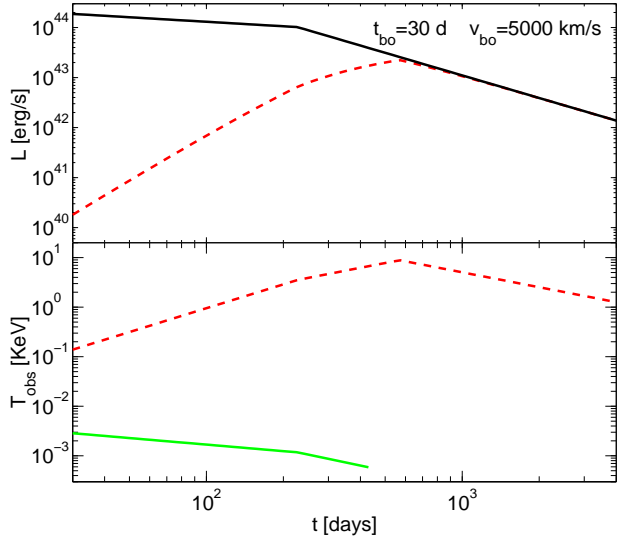}
\caption{An intermediate regime breakout, of $t_{bo}=30$ d and
$v_{bo}=5,000$ $\rm{km\,s^{-1}}$. $t_{SP}$ is calculated for
$E_{51}=1$ and $M_{10}=1$ and $T_{soft}$ calculated for
$\kappa_{bf}\leq \kappa_{ff}$. The transition to the snowplow phase
happens at day $\approx 230$, while the hard component gains
dominance only after this transition, reaching a lower luminosity
and softer X-ray temperatures compared to those of the early regime.
If the breakout is early enough, so that $t_{\rm{hard}} \sim
t_{SP}$, such events produce the brightest X-ray luminosity among
all the cases. Notations are similar to figures \ref{fig:LateBO} and
\ref{fig:EarlyBO}.}\label{fig:IntermidiateBO}
\end{figure}
Figure \ref{fig:IntermidiateBO} presents the luminosity and
temperature of an intermediate case, where the breakout takes place
before the transition to the snowplow phase, but the hard component
may become dominant only after this transition. The properties of
the breakout and the emission that follows soon after it are similar
to those described in the early breakout case. The main difference
from the early breakout case is that the hard component may not
dominate the luminosity at $t_{SP}$ and by the time that it does,
the luminosity may already be dropping fast. If the breakout is
early enough and $t_{\rm{hard}} \sim t_{SP}$, then the X-ray
luminosity must be high. In fact these events produce the brightest
X-ray luminosity among all the cases discussed in this paper.

\section{Summary}
We examine the emission from a shock breakout through a thick wind.
We restrict our exploration to breakout time scales that are days or
longer, which are easier to detect, and velocities in the range
3,000-15,000 $\rm{km\,s^{-1}}$, as expected for typical SNe.
We further restrict our focus to cases where thermal
equilibrium prevails during the evolution of the collisionless
shock. For low metallicities this restriction reduces the shock
velocities of interest to the lower part of the above range. We
consider a standard wind profile $\rho_w \propto r^{-2}$, that is
extended beyond the breakout radius. Following the breakout, the
internal energy in these systems is dominated by the wind material
that is heated by the forward shock
\citep{Chevalier_Irwin2011,Balberg_Loab2011}, which makes a
transition from a radiation mediated shock to a collisionless shock
at the breakout
\citep{Katz+2011}. Our main conclusions are listed below.\\

(i) The breakout and following emission are very luminous. In the
regime that we explore, the shocked plasma is always cooling fast
(see the exact condition in section \ref{sec:tslow}), converting all
the shock luminosity into radiation. Therefore, if the wind mass is
comparable to, or larger than, the ejecta mass, then all the SN
kinetic energy is radiated away. If the wind mass, $M_w$ is smaller
than $M_{ej}$ then the energy radiated during the interaction is
$\sim E M_w/M_{ej}$. This result supports the suggestion by
\cite{Chevalier_Irwin2011} that at
least some ultra-luminous type IIn SNe are such breakouts.\\
(ii) As long as photons that diffuse ahead of the shock
maintain thermal equilibrium with the unshocked gas, the
post-breakout emission is composed of two spectral components - soft
and hard. The electrons behind the collisionless shock are heated to
temperatures of $\sim \min\left[60,200v_9^2\right]$ keV
\citep{Katz+2011}. The soft component is generated by the unshocked
gas ahead of the shock, while the hard component is
generated by the hot shocked electrons via free-free emission and
IC of a small fraction of the soft photons.\\
(iii) The soft component may or may not be in thermal
equilibrium during the breakout, even in the narrow velocity range
of 3,000-15,000 $\rm{km\,s^{-1}}$. Thermal equilibrium is expected
for all slower breakouts $\lesssim 7,000$ $\rm{km\,s^{-1}}$, and
also for faster breakouts if there is a significant bound-free
absorption (i.e., high metallicity). The breakout is then bright in
optical/UV, with a temperature of $10^4-10^5$ K. In faster shocks
that are out of thermal equilibrium the breakout temperature can be
as high as $5 \times 10^6$ K. In that case, the breakout X-rays can
be much brighter than the optical/UV.\\
(iv) If thermal equilibrium is maintained ahead of the shock the
hard component is suppressed as long as IC cooling is more efficient
than free-free cooling \citep{Chevalier_Irwin2012}. In that case
hard photons from free-free emission of hot electrons behind the
forward shock carry at the breakout only $\sim 10^{-4}$ of the soft
component energy. The luminosity of hard photons from IC over the
same electrons may be higher, but it is also a small fraction of the
bolometric luminosity. The hard component luminosity rises quickly
after breakout. It becomes dominant at $\sim 10-50 t_{bo}$.\\
(v) The temperature of the hard component is $m_e c^2/\tau^2$, which
at breakout is $0.1-1$ keV. It rises as $t^2$ after breakouts with
$t_{bo} \lesssim 80$ d and as $t$ for longer breakouts.

The post-breakout evolution can be divided into three regimes
according to the breakout time. (i) Late breakouts (typically
$t_{bo}>80$ d) are very bright in optical/UV and for typical
parameters are very dim in X-rays at early time. $\sim 1$ keV X-rays
peak around $20 t_{bo}$, carrying only $\sim 10^{-2}$ of the
breakout energy. (ii) Early breakout (typically $1<t_{bo}<20$ d)
bolometric luminosity is lower and it can be either in or out of
thermal equilibrium for typical parameters. If it is out of thermal
equilibrium the emission can be bright in X-rays soon after the
breakout (and possibly dim in optical/UV). If it is in thermal
equilibrium then X-rays are suppressed after the breakout, but their
luminosity and hardness rises quickly to gain dominance, at most at
$t \sim 20 t_{bo}$. (iii) Intermediate breakouts (typically $20
{\rm\,d}<t_{bo}<80{\rm\,d}$) are typically in thermal equilibrium
and are therefore bright in optical/UV while X-ray emission is
suppressed at breakout. But, $\sim 1-10$ keV X-rays become dominant
at most at $t \sim 20 t_{bo}$ with a luminosity that may be almost
as high as that of the breakout pulse.

In our calculations we ignore the radius at which the thick wind
ends. Once the wind density drops abruptly the forward shock becomes
inefficient and the luminosity fades quickly
\citep{Chevalier_Irwin2011}. Since the mass of the progenitor at
birth is limited, and the mass in the wind needed for an extended
breakout is considerable, it is expected that the light curves we
present will be terminated at some point. For example, in order for
the light curve to remain bright until $t_{SP}$ the wind mass must
be $\gtrsim M_{ej}$. Even when $M_w > M_{ej}$ the collected mass is
$\propto t^{1/2}$ after $t_{SP}$, limiting the lifetime of the
bright emission. In terms of the prospects of observing X-rays, this
consideration makes earlier breakouts more attractive, since the
mass that early breakout events collect by the time that the hard
component dominates is lower.

We do not address the extinction of soft
X-rays due to photoabsorption by partially ionized unshocked wind \citep{Chevalier_Irwin2012}.
This process may suppress softer-end photons of the hard component
as well as hard photons of the soft component, if the latter is far from thermal equilibrium.
If photoabsorption is important then the observed flux of soft X-rays is lower than the
one that we predict.

Finally, our treatment ignores the contribution of emission from
ejecta mass before it is shocked by the reverse shock. Once the
optical depth of the wind drops, this emission becomes similar to
the emission from a typical SN without a thick wind (e.g., typical
IIP, Ib and Ic SNe). The emission from the forward shock that we
consider here always outshines this emission (in terms of bolometric
luminosity). But, at late times, when most of the forward shock
emission is in hard photons, the contribution from inner layers may
dominate the IR-optical emission.

To conclude, breakouts through thick winds produce very bright SNe
with a potential for detection across the entire electromagnetic
spectrum. The brightest phase is the breakout, which is
typically dominated by optical-UV emission. In that case early X-ray
emission is suppressed, but X-rays becomes harder and their
luminosity rises quickly after the breakout, peaking at most after
$\sim 10-50 \,t_{bo}$.

We thank Eran Ofek and Boaz Katz for helpful discussions and
comments, and the anonymous referee for further helpful comments.
G.S. and E.N. were partially supported by an ISF grant (174/08) and
by an ERC starting grant (GRB-SN 279369). R.S. was partially
supported by ERC and IRG grants, and a Packard Fellowships.

\bibliographystyle{apj}
\bibliography{wind_breakout}

\end{document}